# Some contributions to the theory of edge waves

by R. S. JOHNSON

School of Mathematics & Statistics, University of Newcastle upon Tyne,
Newcastle upon Tyne NE1 7RU, UK

The classical edge-wave problem is addressed by scaling the governing equations, for small slope ($\varepsilon$) at the shore, according to the exact edge-wave solution (for uniform slope) which is based on the Gerstner solution of the water-wave problem. The bottom is allowed to be any suitable profile which varies on the scale of this small parameter; a multiple-scale method is then employed to construct the solution. The leading-order equations – which are a version of the shallow-water equations – are fully nonlinear, but an appropriate exact travelling-wave solution exists; the next term in the asymptotic expansion, valid for $\varepsilon \to 0$, is also found and, from this, uniformity conditions are deduced. The results are used to describe the run-up pattern produced by edge waves at the shoreline, based on any mode other than the first; this pattern corresponds closely with what is observed, and also with the exact solution for uniform slope everywhere. The surface wave, from the shoreline, seawards, is described for various depth profiles (such as a constant depth at infinity or with a sand bar close inshore). The problem for the first mode, which corresponds to a non-uniformity in the expansion, is briefly discussed; in this case it is not possible to find an exact closed-form solution.

The corresponding analysis in the case when a longshore current (varying on the same scale as the depth) is flowing, in addition to a general depth profile, is also presented, and the notion of an effective depth profile is confirmed. Finally, a brief mention is made of model equations for edge waves (which have single-mode exact solutions); these may provide the basis for further investigations into the interaction of modes.

## 1. Introduction

The edge-wave solution of the linear water-wave problem for propagation over a uniform sloping beach was first reported by Stokes (1846). These waves propagate *along* the beach (i.e. in the *longshore* direction) and have an amplitude which decays exponentially away from the shoreline (so they provide an example of a trapped wave); a sloping beach is essential for their existence. Although such waves may be thought to be no more than a mathematical curiosity, they have been found to play an important rôle in many processes near the shoreline; see Howd, Bowen & Holman (1992) (and the many papers cited therein) for an informative background to this problem. In addition, their presence often results in an intriguing – and quite delightful – pattern of trochoidal or, sometimes, cycloidal waves marking the shoreline; some splendid photographs of these run-up patterns can be found in Guza & Inman (1975) and in Komar (1998).

Many authors have made important contributions to the theory of edge waves; we mention a few that have been particularly relevant to the current investigation (and each contains further references for those who may wish to explore the background more deeply). Ursell (1952) demonstrated that more (linear) edge-wave modes appear as the (constant) slope of the beach is progressively decreased; an instructive overview of much of the classical linear theory is presented in Stoker (1957) (and a slightly more modern approach, for simple linear edge waves, is offered in Johnson, 1997). Nonlinear effects in the small-amplitude approximation, based on the shallow-water equations, and then on the full equations (both for constant beach angles), were described by Whitham (1976); this work was extended to more general depth variations by Minzoni (1976). The problem of linear edge waves over a gently sloping beach has been addressed by Miles (1989), in which a result is obtained (for a uniformly valid approximation for the dominant mode) which is generalised here. Mechanisms for the excitation of edge waves have been discussed by Minzoni & Whitham (1977) and by Evans (1988, 1989). Some quite extensive experimental results are reported in Yeh (1985), which describe the evolution and modulation of the edge waves in far more detail than we attempt here. However, the main impetus for the work that we present came from another quarter.

In 1966, Yih demonstrated that a coordinate transformation of Gerstner's 1802 exact, non-trivial solution of the classical water-wave problem for infinite depth (see Lamb, 1932), produces an edge-wave solution for a beach of constant slope. A similar observation is given by Mollo-Christensen (1982). These authors provide an implicit solution of the problem, but this has recently been improved by Constantin (2001), and on two levels. First, he proved that the flow was dynamically possible – the flow map is a diffeomorphism – and then he produced a particularly simple parametric representation of the run-up pattern on the beach (by using the Lagrangian description of the flow field). It was this version of the exact solution which prompted the investigation that we describe here. In essence, we will allow the depth to vary on a slow scale (which will be the slope at the beach), introduce scalings consistent with the Gerstner-Constantin solution and then develop an asymptotic solution by employing the method of multiple scales. By virtue (we might argue) of the existence of the exact solution for a uniform slope, it is possible to find an exact solution for variable depth, at leading order, without the need to invoke an additional small-amplitude approximation. However, because of the particular form of solution that we seek here, we generate a solution which is irrotational, whereas the original Gerstner solution, and Constantin's, describes a solution with a non-zero vorticity. (A small-amplitude approximation can always be introduced if that might lead to useful additional results.)

On following the formulation of the problem, based on suitable scaled variables defined in terms of the small slope ($\varepsilon$) of the beach, the bottom profile is allowed to be a general function of the appropriate 'slow' variable. The first two terms in the asymptotic solution, uniformly valid in space (and time) as $\varepsilon \to 0$, are obtained and a representation of the surface wave is given; in particular, the run-up pattern is described. The properties of the solution for various depth profiles are also presented, including the cases of a constant depth at infinity or with a sand bar fairly close inshore. We also include, in addition to a general depth variation, a longshore current (which varies on the same scale as the depth). We confirm that some elements of the solution are described by replacing the actual depth profile by an 'effective' one, which incorporates the given longshore current; this aspect of our new profile agrees with that introduced by Howd, *et al*. (1992). Finally, some model equations with exact solutions, which represent a single mode, are offered; these equations may prove worthy of some further investigation, particularly as a means for exploring the nature of the interactions between two or more modes.

## 2. Governing equations

We consider an incompressible, inviscid fluid which is bounded above by a free surface ($z = h(x, y, t)$) and below by a fixed, impermeable bed ($z = b(x)$, so not a function of *y*). In its undisturbed state, the free surface is $z = 0$ and this intersects $z = b(x)$ at $x = 0$ – the shoreline in the absence of waves; the fluid extends to infinity as $x \to \infty$ and otherwise $-\infty < y < \infty$ (where *y* is the longshore coordinate). This configuration is shown in figure 1. We choose to use a typical (or mean) wave length, $\lambda$, of the edge waves as the length scale and $\sqrt{g\lambda}$ as the speed scale; thus we take $\lambda/\sqrt{g\lambda}$ as the appropriate time scale. The pressure ($P$) is written as

$$P = P_a - \rho g z + \rho g \lambda p$$

where $\rho$ is the constant density of the water, $g$ is the constant acceleration of gravity and $P = P_a = \text{constant}$ is the pressure at the free surface. (We ignore the rôle of surface tension in this model.) The Euler equation, equation of mass conservation and the boundary conditions, written in non-dimensional variables, are then

$$\frac{D\mathbf{u}}{Dt} = -\nabla p; \quad \nabla \cdot \mathbf{u} = 0 \quad (\mathbf{x} \equiv (x,y,z), \ \mathbf{u} \equiv (u,v,w))$$

with
$$p = h \ \& \ w = \frac{Dh}{Dt} \quad \text{on} \ z = h(x,y,t)$$

and
$$w = u\frac{db}{dx} \quad \text{on} \ z = b(x),$$

where
$$\frac{D}{Dt} \equiv \frac{\partial}{\partial t} + u\frac{\partial}{\partial x} + v\frac{\partial}{\partial y} + w\frac{\partial}{\partial z}.$$

It will be assumed that suitable initial data exist which will generate the edge waves. We use the Euler equation as the basis for this formulation because we make no assumptions, *ab initio*, about the rotationality of the flow field; see Constantin (2001). We will find, in the event, that our construction produces a solution which is irrotational.

The edge waves propagate parallel to the shore (so in the *y*-direction) and otherwise they have a structure in the *x*-direction; in particular, the wave amplitude decays exponentially away from the shoreline. It is likely that the pattern observed at the shoreline is generated by standing waves, but we shall follow the conventional route in this discussion and analyse only travelling waves. In order to accommodate this configuration, and to be consistent with the appearance of the slope of the uniform bed in Constantin (2001), we introduce suitable scaled variables. Let $\varepsilon$ be the magnitude of the slope $b'(0)$ and, further, we assume that the depth varies slowly on this scale; we define the bottom by $z = b = -B(X)$, $X = \varepsilon x$, with $B'(0) = 1$. (The change of sign is merely a convenience.) In addition, we define

$$\xi = \ell y - \omega\sqrt{\varepsilon}\, t, \quad \theta = \varepsilon^{-1} \int_0^X \alpha(X'; \varepsilon)\, dX',$$

where $\ell\ (>0)$ is a given wave number and $\omega$ (= constant) is to be determined, as is $\alpha(X; \varepsilon)$. In Constantin (2001), the velocity components, $(u, v, w)$, are easily seen to be proportional, correspondingly, to $\left(\cos\alpha\sqrt{\sin\alpha}, \sqrt{\sin\alpha}, \sin\alpha\sqrt{\sin\alpha}\right)$, where $\alpha$ is the uniform slope of the bottom; similarly, $p$ and $h$ are proportional to $\sin\alpha$. Thus in our formulation, **u**, $p$ and $h$ are rescaled according to

$$(u, v, w) \to \sqrt{\varepsilon}(u, v, \varepsilon w); \quad (p, h) \to \varepsilon(p, h);$$

the resulting non-dimensional, scaled equations are

$$\frac{Du}{Dt} = -(\alpha p_\theta + \varepsilon p_X); \quad \frac{Dv}{Dt} = -\ell p_\xi; \quad \varepsilon \frac{Dw}{Dt} = -p_z; \qquad (1\text{a,b,c})$$

$$\alpha u_\theta + \ell v_\xi + \varepsilon u_X + \varepsilon w_z = 0, \qquad (2)$$

with $\quad p = h \ \& \ w = -\omega h_\xi + \alpha u h_\theta + \ell v h_\xi + \varepsilon u h_X \ \text{on} \ z = \varepsilon h \qquad (3\text{a,b})$

and $\quad w = -u B'(X) \ \text{on} \ z = -B(X), \qquad (4)$

where $\quad \dfrac{D}{Dt} \equiv -\omega\dfrac{\partial}{\partial\xi} + \alpha u\dfrac{\partial}{\partial\theta} + \ell v\dfrac{\partial}{\partial\xi} + \varepsilon u\dfrac{\partial}{\partial X} + \varepsilon w\dfrac{\partial}{\partial z}.$

(We have used subscripts, where convenient, to represent partial derivatives.) These equations, (1) – (4), provide the basis for our discussion of the problem, in the limit $\varepsilon \to 0$. Because these equations have been scaled for small slope, they are a version of the shallow-water equations. It should be noted, however, that this particular scaling has resulted in the nonlinearity being retained at leading order; if a small-amplitude approximation is also of interest, then the further scaling

$$(u, v, w, p, h) \to \delta(u, v, w, p, h), \text{ with } \delta \to 0, \qquad (5)$$

can be adopted.

## 3. Asymptotic procedure and basic results

Each of the variables $(u,v,w,p,h)$ is written as an asymptotic expansion in $\varepsilon$:

$$q \sim \sum_{n=0}^{\infty} \varepsilon^n q_n \quad (q \equiv u,v,w,p,h)$$

and the problem at each order is formulated. It is expected that uniformity conditions will need to be imposed, presumably as $|\theta| \to \infty$, and that these will be necessary in order to determine completely the earlier terms in the expansions. Furthermore, we observe that the velocity component in the $z$-direction, $w$, appears at $O(\varepsilon)$ relative to the other components $(u, v)$ in equation (2); it is usual, in shallow-water approximations, to find that $w$ is determined at leading order, rather than successively from terms *lower* down the expansion, as is the case here. In addition to the expansion of the dependent variables, we also allow

$$\alpha(X;\varepsilon) \sim \alpha_0(X) + \sum_{n=2}^{\infty} \varepsilon^n \alpha_n(X), \tag{6}$$

where the term $\varepsilon \alpha_1(X)$ is omitted because it can be subsumed into any amplitude function (which, in general, depends on $X$); there is no advantage – at least, to the order we work to here – in also expanding the constant $\omega$.

The leading-order problem, obtained from equations (1a), (1b), (1c), (2) and (3a), respectively, is

$$-\omega u_{0\xi} + \alpha_0 u_0 u_{0\theta} + \ell v_0 u_{0\xi} = -\alpha_0 p_{0\theta}; \quad -\omega v_{0\xi} + \alpha_0 u_0 v_{0\theta} + \ell v_0 v_{0\xi} = -\ell p_{0\xi};$$

$$p_{0z} = 0 \quad \text{and} \quad \alpha_0 u_{0\theta} + \ell v_{0\xi} = 0,$$

with $\qquad\qquad\qquad\qquad p_0 = h_0 \quad \text{on} \quad z = 0.$

This set of shallow-water equations has the particular exact solution (selected by the requirement for the velocity components to be simple trigonometric functions of $\xi$, a form that corresponds to the irrotational solution of Stokes, and others)

$$p_0 = h_0 = A_0(X) e^\theta \cos\xi - \frac{1}{2} \frac{\ell^2}{\omega^2} A_0^2 e^{2\theta}; \tag{7a}$$

$$u_0 = -\frac{\ell}{\omega} A_0 e^{\theta} \sin\xi; \quad v_0 = \frac{\ell}{\omega} A_0 e^{\theta} \cos\xi, \qquad (7b,c)$$

for arbitrary $A_0(X)$ and $\omega$; we have chosen to set $\alpha_0 = -\ell$ so that $\theta \sim -\ell X/\varepsilon$ (i.e. $e^{\theta} \to 0$ as $X \to +\infty$). This solution, which appears to be a new exact solution of the shallow-water equations which does not exhibit wave steepening, is then consistent with the initial surface profile

$$h \sim A_0(\varepsilon x) e^{-\ell x} \cos(\ell y) - \frac{\ell^2}{2\omega^2} A_0^2 e^{-2\ell x}$$

for some $A_0(\varepsilon x)$.

At the next order, we obtain the set of equations

$$-\omega u_{1\xi} + \alpha_0 (u_0 u_1)_\theta + \ell(v_0 u_{1\xi} + v_1 u_{0\xi}) + u_0 u_{0X} = -(\alpha_0 p_{1\theta} + p_{0X}); \qquad (8)$$

$$-\omega v_{1\xi} + \alpha_0 (u_0 v_{1\theta} + u_1 v_{0\theta}) + \ell(v_0 v_1)_\xi + u_0 v_{0X} = -\ell p_{1\xi}; \qquad (9)$$

$$-\omega w_{0\xi} + \alpha_0 u_0 w_{0\theta} + \ell v_0 w_{0\xi} = -p_{1z}; \quad \alpha_0 u_{1\theta} + \ell v_{1\xi} + u_{0X} + w_{0z} = 0, \quad (10,11)$$

with $\qquad\qquad p_1 = h_1 \ \& \ w_0 = -\omega h_{0\xi} + \alpha_0 u_0 h_{0\theta} + \ell v_0 h_{0\xi} \ \text{on} \ z = 0 \qquad (12a,b)$

and $\qquad\qquad\qquad w_0 = -u_0 B'(X) \ \text{on} \ z = -B(X). \qquad (13)$

(The boundary conditions on the surface, $z = \varepsilon h$, are rewritten to all orders as evaluations on $z = 0$ by assuming the existence of Taylor expansions about $z = 0$.) This set has the solution for $v_1$, with $z \in [-B(X), 0]$, which takes the form

$$v_1 = \ell A_{1\xi} + V_1,$$

where $A_1(\xi, \theta, X)$ satisfies

$$\ell^2 \left(A_{1\theta\theta} + A_{1\xi\xi}\right) + \frac{1}{\omega B}\left(\omega^2 A_0 - \ell A_0 B' - 2\ell A_0' B\right) e^{\theta} \sin\xi - \frac{\ell^4 A_0^3}{\omega^3 B} e^{3\theta} \sin\xi = 0;$$

the function $V_1$, which is bounded, is defined below. (Corresponding expressions appear for $u_1$ and $h_1$.) Now the asymptotic expansion for $v$ (and, similarly, for $u$ and $h$) is uniformly valid as $\theta \to -\infty$ and $|\xi| \to \infty$ only if the coefficient of the term $e^{\theta} \sin\xi$ in the equation for $A_1$ is zero i.e.

$$A_0 B' + 2 B A_0' = \frac{\omega^2}{\ell} A_0.$$

With this condition imposed, the solution of the set can be written as

$$w_0 = \left\{\omega - \frac{\ell^4}{\omega^3} A_0^2 e^{2\theta} + \frac{1}{B}\left(\omega + \frac{\ell}{\omega} B' - \frac{\ell^4}{\omega^3} A_0^2 e^{2\theta}\right) z\right\} A_0 e^{\theta} \sin\xi + W_0; \quad (14)$$

$$u_1 = \left(\frac{A_0'}{\omega} - \frac{3\ell^3}{8\omega^3} \frac{A_0^3}{B} e^{2\theta}\right) e^{\theta} \sin\xi + U_1; \quad v_1 = \frac{\ell^3}{8\omega^3} \frac{A_0^3}{B} e^{3\theta} \cos\xi + V_1, \quad (15,16)$$

where the functions $W_0$, $U_1$, $V_1$ and $p_1$ are described in the Appendix; these do not contribute to the description of the surface perturbation, $h_1$, which becomes

$$h_1 = \frac{\ell^2}{8\omega^2} \frac{A_0^3}{B} e^{3\theta} \cos\xi + \frac{\ell}{8\omega^2}\left(\frac{\ell^3}{\omega^2} \frac{A_0^3}{B} e^{2\theta} - 4 A_0'\right) A_0 e^{2\theta} \cos 2\xi$$

$$+ \frac{\ell}{4\omega^2}\left(2 A_0' - \frac{\ell^3}{\omega^2} \frac{A_0^3}{B} e^{2\theta}\right) A_0 e^{2\theta}. \quad (17)$$

The uniformity condition is solved to give

$$A_0(X) = \frac{1}{\sqrt{B(X)}} \exp\left\{\frac{\omega^2}{2\ell} \int^X \frac{dX'}{B(X')}\right\} \quad (18)$$

which is a generalisation of a result obtained by Miles (1989) as a contribution to his uniformly valid expansion for the linear, dominant mode for edge waves propagating over a small, uniform slope. The non-uniformity that is evident in (15) and (16), as $B \to 0$ for general $\omega$, will be addressed below.

The investigation was continued (in outline) as far as the next terms in the expansions, but no additional complications were encountered. A uniformly valid solution can be found, although a similar analysis to that described in the Appendix (for $W_0$, $U_1$, $V_1$ and $p_1$) is required at this order and, indeed, at every order hereafter. The evidence of the first two terms of this type, in conjunction with the general structure of this problem, suggests that no non-uniformities arise from these contributions to the solution. The solutions appearing at higher order merely produce (small) corrections to the solution as presented here, although considerable technical complications are evident in the formulation; because of this, we do not record the details.

The vorticity, **ω**, of this flow field, following the non-dimensionalisation and scaling introduced earlier (with $\boldsymbol{\omega} \to \sqrt{\varepsilon}\,\boldsymbol{\omega}$), can be written

$$\boldsymbol{\omega} \equiv \left(\varepsilon \ell w_\xi - v_z,\; u_z - \varepsilon(\alpha w_\theta + \varepsilon w_X),\; \alpha v_\theta + \varepsilon v_X - \ell u_\xi\right), \tag{19}$$

where $(u, v, w)$ are the velocity components used in equations (1) – (4). When this is calculated for the velocity field

$$u = u_0 + \varepsilon u_1 + \mathrm{O}(\varepsilon^2);\quad v = v_0 + \varepsilon v_1 + \mathrm{O}(\varepsilon^2);\quad w = w_0 + \mathrm{O}(\varepsilon),$$

using (14), (15) and (16) (together with the further details given in the Appendix), we find that $\boldsymbol{\omega} = \mathrm{O}(\varepsilon^2)$, i.e. the flow field is irrotational, correct at $\mathrm{O}(\varepsilon)$. Thus our procedure is generating a solution which is an extension of the classical results discussed by Stokes (1846) and Whitham (1976), for example, but without invoking a small-amplitude approximation. On the other hand, the exact solution for $B(X) = X$, described by Constantin (2001), possesses a non-zero vorticity (which is always in the direction perpendicular to the bottom). This exact solution, following our non-dimensionalisation and scaling, can be expressed as

$$\mathbf{u} \equiv \frac{1}{\sqrt{\ell}} \mathrm{e}^{\ell(b-c)} (\sin\varsigma,\; -\cos\varsigma,\; -\sin\varsigma)$$

where $\quad \varsigma = a\ell - t\sqrt{\varepsilon \ell} \quad$ and $\quad y = a - \ell^{-1} \mathrm{e}^{\ell(b-c)} \sin\varsigma,$

together with corresponding definitions for $x$ and $z$; see Constantin (2001). Here, the parameters $a$, $b$ and $c$ describe the position of a particle at $t = 0$ in a Lagrangian representation (although the identification of that position is not simply $(a,b,c)$). Now, in our construction of the solution, we have taken the longshore propagation variable to be simply $\xi = \ell y - (\omega\sqrt{\varepsilon})t$ i.e. $a = y$ (and $\omega = \sqrt{\ell}$ here). We might surmise, therefore, that if we seek a more general solution in which $\xi$ is defined to include a nonlinear correction, an asymptotic form of the *rotational* solution will be generated; this is left as an investigation for the future.

## 4. The edge wave

At the order to which we have obtained the details, the surface wave is

$$h(\theta, X, \xi; \varepsilon) \sim A_0 e^\theta \cos\xi - \frac{\ell^2}{2\omega^2} A_0^2 e^{2\theta} + \varepsilon \left\{ \frac{\ell^2}{8\omega^2} \frac{A_0^3}{B} e^{3\theta} \cos\xi \right.$$

$$\left. + \frac{\ell}{8\omega^2} \left( \frac{\ell^3}{\omega^2} \frac{A_0^3}{B} e^{2\theta} - 4 A_0' \right) A_0 e^{2\theta} \cos 2\xi + \frac{\ell}{4\omega^2} \left( 2 A_0' - \frac{\ell^3}{\omega^2} \frac{A_0^3}{B} e^{2\theta} \right) A_0 e^{2\theta} \right\} \quad (20)$$

where

$$A_0(X) = \frac{1}{\sqrt{B(X)}} \exp\left\{ \frac{\omega^2}{2\ell} \int^X \frac{dX'}{B(X')} \right\}. \quad (21)$$

The shore (beach) is described by $B(X) \sim X$ as $X \to 0$, and so

$$A_0(X) \sim k X^\beta, \quad \beta = \frac{1}{2}\left( \frac{\omega^2}{\ell} - 1 \right), \text{ as } X \to 0, \quad (22)$$

where $k$ is a constant which is determined by the amplitude of the wave for some $X > 0$. If $A_0(X)$, and all its derivatives, exist as $X \to 0$, then we require

$$\beta = \frac{1}{2}\left( \frac{\omega^2}{\ell} - 1 \right) = n, \quad n = 0, 1, 2, \ldots, \quad (23)$$

which recovers the classical result for the modes of linear edge waves (conventionally obtained *via* the properties of the Laguerre equation). Further, this result also affords a measure of agreement with Minzoni (1976) where variable depth (in particular, finite depth at infinity) was incorporated within the shallow-water model, yet the eigenvalue problem is the same as for the classical constant-slope problem. However, the uniform validity of our asymptotic expansion (20) imposes an additional constraint: we require that $(A_0^2/B)e^{2\theta}$ remains bounded for all $X$, from the run-up on the beach to infinity. For $X \to 0$ (which will apply in the neighbourhood of the run-up), we must have $2n > 1$ (when we elect to use (23)), and so our asymptotic solution is not valid for the lowest mode ($n = 0$), but it does hold for all the others ($n = 1, 2, \ldots$). The exponential decay as $X$ (and $x$) $\to +\infty$ ensures the validity seawards. A discussion of the lowest mode is given in section 5, although – as we indicate there – a leading-order, closed-form solution has not been found in this case.

The run-up pattern on the beach is given by the intersection of the surface wave with the bottom profile there i.e.
$$z = -B(X) = \varepsilon h$$
and with $B(X) \sim X$ as $X \to 0$, we will take this to be

$$-x \sim A_0 e^\theta \cos\xi - \frac{\ell^2}{2\omega^2} A_0^2 e^{2\theta} \qquad (24)$$

(and the correction term, $\varepsilon h_1$ from (17), could be included if that was thought to be useful). For the discussion presented here, we take (24) as the equation that describes the run-up pattern at the shoreline, with $A_0(X)$ given by (22) (for $\beta = n$) and $\theta = -\ell X/\varepsilon = -\ell x$. In this equation, (24), because it has been generated by a multiple-scale technique, we should treat $X$, $\theta$ and $\xi$ each as O(1) and independent; however, for the purposes of producing graphical results, we must be somewhat cavalier in our interpretation. Thus we choose to use a suitable normalised version of this equation:

$$1 + \mu Z^{n-1} e^{-Z} \cos\xi - \frac{\mu^2}{2(1+2n)} Z^{2n-1} e^{-2Z} = 0, \qquad (25)$$

where $Z = \ell x$, $\mu = k\varepsilon^n/\ell^{n-1}$ (for $n = 1, 2, ...$) and the root $Z = 0$ has been eliminated. We suggest, even with $\mu = O(1)$, that this is worth examining as a basis for a representation of the run-up pattern. Indeed, because *k* here is associated with the amplitude of the wave (see (22)), some freedom in its choice is permitted (although, formally, for *k* and $\ell$ both O(1), we see that $\mu = O(\varepsilon^n)$). It is quite straightforward to show that solutions exist of this equation that are continuous, bounded and periodic for $\mu \geq \mu_n > 0$ (for suitable $\mu_n$, described below). The solutions for $\mu < \mu_n$ comprise closed curves, spaced periodically, which coalesce for $\mu = \mu_n$ to form two near-cycloids that meet at their cusps; for $\mu > \mu_n$, these curves separate to become a pair of curves that correspond to trochoids. These profiles are analogous to the cycloid and trochoid given in Constantin (2001), although here we have a pair in each case, and either is an acceptable solution – pointing either towards the shore or seawards. (That these two possibilities can occur appears to be borne out in some of the observations of edge waves i.e. profiles can 'point' either seawards, or towards the beach; see Guza & Inman, 1975, and Komar, 1998.) For equation (25), a routine numerical investigation yields the values $\mu_1 \approx 7 \cdot 27$, $\mu_2 \approx 5 \cdot 87$, $\mu_3 \approx 2 \cdot 67$; an example of a cycloid-like profile ($n = 2$, $\mu = 5 \cdot 865$) is presented in figure 2a (this being one of a pair), and a pair of trochoid-like profiles ($n = 1$, $\mu = 8$) is shown in figure 2b. By comparison, the exact solution (Constantin, 2001) requires, of course, the choice $B(X) = X$ (for $\forall X$) and, in addition, this solution corresponds only to the lowest, linear mode (equivalently $n = 0$ here); the parameter $b_0 \leq 0$ in Constantin plays the rôle of our $\mu \geq \mu_n$. Even though we have been cavalier with our interpretation of $\mu$, we submit that the model run-up pattern, obtained from (25), based on a slowly varying depth, has successfully captured all the essential features of this phenomenon, albeit excluding the first mode ($n = 0$). However, the solutions with closed curves ($\mu < \mu_n$) cannot be simply interpreted. That they correspond, for example, to solutions with 'holes' – where the bottom is uncovered – is unlikely because, for $\mu \geq \mu_n$, the ocean extends no further than either one or the other boundary curve i.e. never beyond the one furthest inshore. For a hole to appear, the water would need to extend beyond this furthest boundary (and so exist indefinitely up the shoreline). We suggest, at this stage of the investigation, that solutions for $\mu < \mu_n$ do not describe a physically realistic phenomenon.

These edge waves have the familiar structure of a trapped wave, which is evident here by virtue of the exponential decay (terms $e^{m\theta}$, $m = 1, 2, 3$, in equations (7) and (14) – (17)). In particular, the amplitude function (which predominantly controls the form of the surface wave), as a function of $x$, is

$$A_0(X) e^\theta = \frac{1}{\sqrt{B(\varepsilon x)}} \exp\left\{ \frac{\omega^2}{2\ell} \int^{\varepsilon x} \frac{dX}{B(X)} - \ell x \right\} \qquad (26)$$

and this allows a detailed investigation of the effects of various depth profiles. This amplitude in the presence of the uniform slope, $B(X) = X$, is proportional to

$$Z^n e^{-Z} \quad (Z = \ell x, \ \omega^2/2\ell = n + \tfrac{1}{2}) \qquad (27)$$

which we will need for the purposes of comparison. If the depth is finite at infinity ($X \to +\infty$), then we may choose to model the bottom profile by

$$B(X) = B_\infty\left(1 - e^{-X/B_\infty}\right),$$

which satisfies the given condition $B'(0) = 1$ at the shoreline; the amplitude is then proportional to

$$\lambda^{-n}\left(e^{\lambda Z} - 1\right)^n \exp\left\{-\left(1 - \tfrac{1}{2}\lambda\right)Z\right\}, \quad \lambda = \varepsilon/(\ell B_\infty), \tag{28}$$

where we require $\lambda < (n + \tfrac{1}{2})^{-1}$ in order to maintain the exponential decay at infinity. This expression, (28), has been chosen so that it recovers (27) for $\lambda \to 0$ (at fixed $Z$), enabling us directly to compare the results. The amplitude function from the shoreline, seawards, for (27) and also some choices of $\lambda$ in (28), is shown in figure 3 (all for $n = 2$). The effect of allowing finite depth at infinity is, not surprisingly, to increase the maximum value of the amplitude function.

A more interesting example is provided by the case of a sand bar fairly close inshore, in a depth profile that is otherwise linear. Some amplitude profiles, all of which correspond to (27) as $Z \to 0$, are shown in figure 4. The effect of a sand bar is quite dramatic: there is a significant increase in the amplitude just behind (seawards) of the bar (as alluded to in Howd, *et al.*, 1992; see also Kirby, Dalrymple & Liu, 1981). Of course, any suitable depth profile can be chosen – perhaps a sand bar combined with finite depth at infinity – and its effects investigated.

Finally, we collect these ideas and so present the leading term for the edge waves; with the same notation as we used in (25), the surface wave is proportional to

$$Z^n e^{-Z} \cos\xi - \frac{\mu}{2(1 + 2n)} Z^{2n} e^{-2Z} \tag{29}$$

in the case $B(X) = X$. An example of the surface profile ($n = 4$, $\mu = 4$), with its run-up pattern at the shoreline, is shown in figure 5.

## 5. The first mode ($n = 0$)

The solution that we have presented for the edge waves is uniformly valid only if $n = 1, 2, \ldots$, thereby excluding the first mode. In the case $n = 0$, with $B(X) \sim X$ as $X \to 0$, our expansions are not valid as $X \to 0$: they break down where $X = O(\varepsilon)$ i.e.

$x = \mathrm{O}(1)$. In this region we seek a solution for $\varepsilon h \geq z \geq X = \varepsilon x$, and so we introduce $z = \varepsilon Z$ and use $x$ (rather than $X$). The leading-order problem, as $\varepsilon \to 0$, is then described by the equations

$$-\omega u_\xi + \alpha u u_\theta + \ell v u_\xi + u u_x + w u_Z = -(\alpha p_\theta + p_x);$$

$$-\omega v_\xi + \alpha u v_\theta + \ell v v_\xi + u v_x + w v_Z = -\ell p_\xi;$$

$$p_Z = 0; \quad \alpha u_\theta + \ell v_\xi + u_x + w_Z = 0,$$

with $p = h$ & $w = -\omega h_\xi + \alpha u h_\theta + \ell v h_\xi + u h_x$ on $Z = h$ and $w = -u$ on $Z = -x$.

The exact solution of these equations, relevant in this context, has not been found, but it is a routine exercise to show that a suitable expansion of the solution confirms that matching is possible. In particular, when we seek a solution

$$\phi \sim \sum_{n=1}^{\infty} F_n(\xi, x, Z) \mathrm{e}^{n\theta} \text{ and } h \sim \sum_{n=1}^{\infty} H_n(\xi, x) \mathrm{e}^{n\theta},$$

where $\phi$ is the velocity potential ($u = \alpha \phi_\theta + \phi_x$, $v = \ell \phi_\xi$), and solve at each order in $\mathrm{e}^{n\theta}$, we find that

$$h \sim A \mathrm{e}^\theta \cos \xi - \frac{\ell^2}{2\omega^2} A^2 \mathrm{e}^{2\theta} + \frac{\ell^3}{4\omega^2} A^3 \mathrm{e}^{3\theta} \left\{ \mathrm{e}^{2\ell x} \int_x^\infty \frac{\mathrm{e}^{-2\ell y}}{y} \mathrm{d}y \right\} \cos \xi$$

for arbitrary $A$ (constant) and with truncation imposed beyond the term in $\mathrm{e}^{3\theta}$. This solution matches precisely to our expansion for $h$, (20), in the case $n = 0$ with $X \to 0$. If this were to be regarded as a reasonable model for the run-up pattern associated with the first mode then, corresponding to (25), we would have

$$Y + \left\{ \mu \mathrm{e}^{-Y} + \frac{1}{4} \mu^3 \mathrm{e}^{-Y} \int_Y^\infty \frac{\mathrm{e}^{-2y}}{y} \mathrm{d}y \right\} \cos \xi - \frac{1}{2} \mu^2 \mathrm{e}^{-2Y} = 0,$$

with the same definitions as before (but $Y$ here replacing the $Z$ used in (25)). This equation predicts a run-up pattern which can never accommodate cusps. However, without the advantage of an exact solution of our set of equations, we can offer no more at this stage; further investigation (perhaps numerical) is deferred for the present.

## 6. The effects of a longshore current

Longshore currents are known to play an important rôle in the structure of edge waves and, although they are not always present, they do occur often enough to encourage further study; see Kenyon (1972) and Howd, *et al.* (1992). The variation of the longshore current in the seawards direction is controlled, to some extent, by the depth profile and so we assume in this model that it varies on the same scale as the depth. The governing equations are precisely those already presented, in (1) – (4), but with $v(\theta, X, \xi, z; \varepsilon)$ replaced by $V(X) + v(\theta, X, \xi, z; \varepsilon)$, where $V(X)$ is the given longshore current. Note that both $V$ and $v$ are the same size. The development follows very closely that already described, and eventually produces, in place of (20), the surface wave

$$h(\theta, X, \xi; \varepsilon) \sim A_0 e^\theta \cos\xi - \frac{\ell^2 A_0^2 e^{2\theta}}{2(\omega - \ell V)^2} + \varepsilon \left\{ \left[ \frac{U_0 V'}{\ell} e^\theta - \frac{U_0^3}{8B}\left(\frac{\omega}{\ell} - V\right) e^{3\theta} \right] \cos\xi \right.$$

$$\left. + \frac{U_0}{8}\left[ \frac{U_0^3}{B} e^{2\theta} - \frac{4 U_0'}{\ell} \right] e^{2\theta} \cos 2\xi + \frac{U_0}{2}\left( \frac{U_0'}{\ell} - \frac{U_0^3}{2B} e^{2\theta} \right) e^{2\theta} \right\}, \quad (29)$$

where $U_0 = -\ell A_0/(\omega - \ell V)$. The amplitude, $A_0(X)$, is now given by

$$A_0 B' + 2 B A_0' = \frac{(\omega - \ell V)^2}{\ell} A_0 - \frac{2\ell B V'}{\omega - \ell V} A_0$$

which yields (cf. (18))

$$A_0(X) = \frac{(\omega - \ell V)}{\omega \sqrt{B(X)}} \exp\left\{ \frac{1}{2\ell} \int^X \frac{[\omega - \ell V(X')]^2}{B(X')} dX' \right\}. \quad (30)$$

An additional factor of $\omega$ is included here in order to allow direct correspondence with our earlier result, (18); we may elect to write the solution in this form because of the arbitrary multiplicative constant that is associated with the indefinite integral in the exponent.

It is immediately evident that (30) can be expressed in exactly the same form as (18), where the $B(X)$ there is replaced by the 'effective' depth profile

$$\overline{B}(X) = \frac{B(X)}{\left(1 - \frac{\ell V}{\omega}\right)^2} ; \qquad (31)$$

see Howd, *et al*. (1992). We assume that the longshore current satisfies $V(X) < \omega/\ell$ for all choices of $\omega$. (If $V(X)$ approaches $\omega/\ell$ for any particular $X$, then presumably edge waves no longer exist; we do not explore this complication here.) The result of combining a typical longshore current of the form

$$V(X) = X e^{-X/X_0}, \ X \geq 0, \ X_0 > 0,$$

for example, with a uniform depth profile ($B(X) = X$), is to produce an effective profile which can incorporate a sand bar; see figure 6. However, although $A_0$ can be written in terms of $\overline{B}$, the rest of the expression for the surface wave, given in (29), requires explicit use of $V(X)$; so for example, the leading approximation to this wave becomes

$$h \sim \overline{A}_0 e^{\theta} \cos \xi - \frac{\ell^2 \overline{A}_0^2}{2(\omega - \ell V)^2} e^{2\theta}$$

where $\overline{A}_0$ is $A_0$ expressed in terms of $\overline{B}$; this can be used to investigate the effects of suitable choices for $V(X)$.

## 7. Discussion

The classical water-wave problem, as it applies to the propagation of edge waves, has been presented in a form consistent with the recent exact solution given by Constantin (2001), but recast for an arbitrary depth profile that varies on a suitable small scale ($\varepsilon$). The resulting problem, treated as asymptotic for $\varepsilon \to 0$, is fully nonlinear at leading order, but with a relevant exact solution. This solution has been used to give a representation of the run-up pattern at the shoreline, and it would appear that this captures most elements of the patterns that are observed (and are seen in the exact solution for a uniform slope: Constantin, 2001). However, our closed-form results do not apply to the lowest mode (as interpreted by the value of *n* in the context of the linear problem), although it is valid for all the other corresponding modes. We have, therefore, given separately a description of the *n* = 0 problem, with an indication that a solution exists which is consistent with the usual run-up pattern. An additional complication in our theory is that the solution for the run-up pattern ($n \neq 0$) comprises two families, and there is no obvious way to select one rather than the other (other than, possibly, by invoking the initial data for particular edge waves); this aspect of the problem is still being explored. The structure of edge waves, away from the shore, can be described for any suitable choice of depth variation. We have included, as examples, the classical uniform-slope case, constant depth at infinity and a model for a sand bar fairly close inshore. The latter two configurations lead to an increase in the maximum amplitude of the edge waves, as compared with the corresponding uniform-slope solution. Thus our version of the theory of edge waves is offered as a simple way, in the first instance at least, of providing an analytical approach in the study of the effects of any chosen depth profile.

The longshore and cross-shore topography which ensures that edge waves can exist, given suitable initial conditions, sometimes has an associated longshore current. This we have also modelled by allowing the current to vary (seawards) on the same scale as the depth profile. The analysis can be carried through, producing a description which mirrors, in the main, the earlier results, by replacing the depth profile by an effective profile which combines the actual profile with the longshore current. This confirms the results obtained by Howd, *et al*. (1992), although we have been able to provide more details (showing, in particular, that the introduction of an effective depth profile is not sufficient for the complete description of the wave, to leading order). Again, these formulae for the structure of edge waves may be used to give – albeit approximately – a simple analytical form of the waves, from the shore to the open ocean.

With the continuing interests in nonlinear equations that are relevant to wave phenomena, and which have exact solutions, we conclude by offering two versions of an equation that might prove worthy of some further study. If we retain the terms in $\varepsilon$, but consider the case of small-amplitude waves (proportional to $\delta$, say; see (5)), then we obtain

$$\left.\begin{array}{l} \alpha^2 \phi_{\theta\theta} + \ell^2 \phi_{\xi\xi} + \varepsilon\left[2\alpha\phi_{\theta X} + \alpha_X \phi_\theta + B^{-1}\left(\alpha B' \phi_\theta - \omega h_\xi\right)\right] = 0; \\ h - \omega\phi_\xi + \tfrac{1}{2}\delta\left(\alpha^2 \phi_\theta^2 + \ell^2 \phi_\xi^2\right) = 0, \end{array}\right\}$$

where we have retained terms as far as $O(\varepsilon)$ and $O(\delta)$, but neglected terms $O(\varepsilon\delta)$; $\phi$ is the velocity potential. This pair has an exact solution, for any given and suitable $B(X)$ (with $\alpha = -\ell$):

$$\phi = A(X)e^{\theta}\sin\xi; \quad h = \omega A e^{\theta}\cos\xi - \tfrac{1}{2}\delta\ell^2 A^2 e^{2\theta}$$

where

$$A(X) = \frac{1}{\sqrt{B(X)}}\exp\left\{\frac{\omega^2}{2\ell}\int^X \frac{dX'}{B(X')}\right\}.$$

This model contains a coupling between the two functions, $\phi$ and $h$; indeed, we may eliminate $h$ (and then $h$ is given by the second of the pair) to produce

$$\alpha^2 \phi_{\theta\theta} + \ell^2 \phi_{\xi\xi} + \varepsilon\left(2\alpha\phi_{\theta X} + \alpha_X \phi_\theta + \alpha B^{-1}B'\phi_\theta\right)$$
$$-\varepsilon\omega B^{-1}\left\{\omega\phi_{\xi\xi} - \tfrac{1}{2}\delta\left(\alpha^2\phi_\theta^2 + \ell^2\phi_\xi^2\right)_\xi\right\} = 0. \qquad (32)$$

A reduced version of this equation, which retains the nonlinearity, is

$$\alpha^2 \phi_{\theta\theta} + \left(\ell^2 - \varepsilon\omega^2 B^{-1}\right)\phi_{\xi\xi} + \tfrac{1}{2}\varepsilon\delta\omega B^{-1}\left(\alpha^2\phi_\theta^2 + \ell^2\phi_\xi^2\right)_\xi = 0,$$

which has been obtained by taking $\varepsilon(\partial/\partial X) \to 0$, but treating $\varepsilon\delta/B$ and $\varepsilon\omega^2/B$ as fixed. Further, because $X$ now appears, at most, as a parameter in this equation, we are at liberty to write it in a normalised form:

$$\Phi_{\theta\theta} + \left(1 - \frac{\varepsilon\omega^2}{B\ell^2}\right)\Phi_{\varsigma\varsigma} + \left(\Phi_\theta^2 + \Phi_\varsigma^2\right)_\varsigma = 0, \qquad (33)$$

which has the exact solution $\Phi = A(X)e^{\theta}\sin\varsigma$, for arbitrary $A$, if we neglect the term in $\varepsilon$. These two equations, (32) and (33), have exact single-component solutions. We suggest that the search for other exact solutions – certainly those that represent the interaction of different modes – is a worthy exercise (which might have to be initiated by a numerical investigation). If this proves to be unsuccessful, the equations – but particularly (32) – can still be used to construct asymptotic solutions for $\varepsilon \to 0$, because this is likely to be simpler than reverting to the original, full equations, and we can see that the equations contain all the essential ingredients for a description of edge waves. These and related issues are to be examined in the near future.

The author is very pleased to acknowledge his thanks to Prof. Adrian Constantin for bringing this problem to his attention, and for encouraging an asymptotic approach that would complement his analysis. Acknowledgement is also due to the referees whose comments led to a number of useful adjustments to an earlier version of this paper.

## Appendix

The solutions given in equations (14), (15) and (16) are written, for convenience, as

$$w_0 = w_{00} + w_{01}z + W_0; \quad u_1 = u_{10} + U_1; \quad v_1 = v_{10} + V_1$$

where $w_{00}(\theta, X, \xi)$, $w_{01}(\theta, X, \xi)$, $u_{10}(\theta, X, \xi)$ and $v_{10}(\theta, X, \xi)$ are defined by these aforementioned equations. Then

$$p_1 = \left(\omega - \frac{\ell^2}{\omega} A_0 e^\theta \cos\xi\right)\left(zw_{00} + \tfrac{1}{2}z^2 w_{01}\right)_\xi$$

$$- \frac{\ell^2}{\omega}\left(zw_{00} + \tfrac{1}{2}z^2 w_{01}\right)_\theta A_0 e^\theta \sin\xi + h_1 + P_1,$$

where 
$$P_{1z} = \left(\omega - \frac{\ell^2}{\omega} A_0 e^\theta \cos\xi\right)W_{0\xi} - \frac{\ell^2}{\omega}\left(A_0 e^\theta \sin\xi\right)W_{0\theta}$$

with $P_1 = 0$ on $z = 0$. The boundary conditions on $W_0$ have now become $W_0 = 0$ on both $z = 0$ and $z = -B(X)$, which ensures that $W_0$ does not contribute to the surface wave. The resulting equations for $W_0$, $U_1$, $V_1$ and $P_1$ are solved by writing first

$$W_0 = \sum_{n=2}^{\infty}\left(Bz^n + z^{n+1}\right)C_n(\theta, X, \xi)$$

which gives

$$P_1 = \sum_{n=2}^{\infty}\left\{\left(\omega - \frac{\ell^2}{\omega} A_0 e^\theta \cos\xi\right)C_{n\xi} - \frac{\ell^2}{\omega}\left(A_0 e^\theta \sin\xi\right)C_{n\theta}\right\}\left(\frac{Bz^{n+1}}{n+1} + \frac{z^{n+2}}{n+2}\right).$$

Now we expand $U_1$ and $V_1$:

$$U_1 = \sum_{n=1}^{\infty} D_n(\theta, X, \xi) z^n \, ; \quad V_1 = \sum_{n=1}^{\infty} E_n(\theta, X, \xi) z^n$$

and hence we may solve sequentially for the coefficients of these series. We find that

$$D_1 = -\ell\left(\omega - \frac{3\ell^4}{\omega^3} A_0^2 e^{2\theta}\right) A_0 e^{\theta} \sin\xi; \quad E_1 = \ell\left(\omega - \frac{\ell^4}{\omega^3} A_0^2 e^{2\theta}\right) A_0 e^{\theta} \cos\xi;$$

$$D_2 = -\frac{\ell}{2B}\left(\omega - \frac{\ell}{\omega} B' - \frac{3\ell^4}{\omega^3} A_0^2 e^{2\theta}\right) A_0 e^{\theta} \sin\xi;$$

$$E_2 = \frac{\ell}{2B}\left(\omega - \frac{\ell}{\omega} B' - \frac{\ell^4}{\omega^3} A_0^2 e^{2\theta}\right) A_0 e^{\theta} \cos\xi; \quad C_2 = \frac{4\ell^6}{\omega^3 B} A_0^3 e^{3\theta} \sin\xi;$$

$$D_3 = -\tfrac{1}{3}\ell B C_{2\theta}; \quad E_3 = \tfrac{1}{3}\ell B C_{2\xi}; \quad C_3 = \frac{16\ell^6}{3\omega^3} \frac{A_0^3}{B^2} e^{3\theta} \sin\xi,$$

and then

$$D_n = -\frac{\ell}{n}(BC_{n-1} + C_{n-2})_\theta; \quad E_n = \frac{\ell}{n}(BC_{n-1} + C_{n-2})_\xi \quad \text{for } n = 4, 5, 6, \ldots$$

with $\quad BC_{n+1} + C_n = \dfrac{\ell}{n+1}\left(D_{n\theta} - E_{n\xi}\right) \quad$ for $n = 3, 4, 5, \ldots$ .

All these series converge for $-1 \leq z/B \leq 0$, with $X > 0$ and $\forall \xi$, provided that terms such as $\left(A_0^2/B\right)e^{\theta}$ remain bounded for $0 < X < \infty$; see section 4. This is sufficient to confirm the existence of the solution that we have presented earlier.

## REFERENCES

Constantin, A. 2001 Edge waves along a sloping beach. *J. Phys. A: Math. Gen.* **34**, 9723-9731.

Evans, D.V. 1988 Mechanisms for the generation of edge waves over a sloping beach. *J. Fluid Mech.* **186**, 379-391.

\_\_\_\_\_\_ 1989 Edge waves over a sloping beach. *Q. J. Mech. Appl. Math.* **42**, 131-142.

Guza, R.T. & Inman, D.L. 1975 Edge waves and beach cusps. *J. Geophys. Res.* **80**, 2997-

3012.

Howd, P.A., Bowen, A.J. & Holman, R.A. 1992 Edge waves in the presence of strong longshore currents. *J. Geophys. Res.* **97**, 11357-11371.

Johnson, R.S. 1997 *A Modern Introduction to the Mathematical Theory of Water Waves*. Cambridge University Press; Cambridge, UK.

Kenyon, K.E. 1972 Edge waves with current shear. *J. Geophys. Res.* **77**, 6599-6603.

Kirby, J.T., Dalrymple, R.A. & Liu, P.L.-F. 1981 Modification of edge waves by barred beach topography. *Coastal Eng.* **5**, 35-49.

Komar, P. 1998 *Beach Processes and Sedimentation*. Prentice-Hall; Englewood Cliffs, NJ.

Lamb, H. 1932 *Hydrodynamics* ($6^{th}$ ed). Cambridge University Press; Cambridge, UK.

Miles, J.W. 1989 Edge waves on a gently sloping beach. *J. Fluid Mech.* **199**, 125-131.

Minzoni, A.A. 1976 Nonlinear edge waves and shallow-water theory. *J. Fluid Mech.* **74**, 369-374.

Minzoni, A.A. & Whitham, G.B. 1977 On the excitation of edge waves on beaches. *J. Fluid Mech.* **79**, 273-287.

Mollo-Christensen, E. 1982 Allowable discontinuities in a Gerstner wave. *Phys. Fluids* **25**, 586-587.

Stoker, J.J. 1957 *Water Waves*. Wiley-Interscience; NewYork.

Stokes, G.G. 1846 Report on recent researches in hydrodynamics. *Rep. $16^{th}$ Brit. Assoc. Adv. Sci.* 1-20. (See also *Papers,* Vol. **I**, 157-187. Cambridge University Press; Cambridge UK, 1880.)

Ursell, F. 1952 Edge waves on a sloping beach. *Proc. R. Soc. Lond.* **A214**, 79-97.

Whitham, G.B. 1976 Nonlinear effects in edge waves. *J. Fluid Mech.* **74**, 353-368.

Yeh, H.H. 1985 Nonlinear progressive edge waves: their stability and evolution. *J. Fluid Mech.* **152**, 479-499.

Yih, C.-S. 1966 Note on edge waves in a stratified fluid. *J. Fluid Mech.* **24**, 765-767.

# Figure captions for *'Some contributions to the theory of edge waves'*

**Figure 1:** Sketch of the surface wave, $z = h(x,y,t)$, and the variable bottom, $z = b(x)$; $y$ is the longshore coordinate and the seawards direction is $x \to \infty$.

**Figure 2: (a)** Cycloid-like run-up pattern, obtained from equation (25), for $n = 2$, $\mu = 5 \cdot 865$. **(b)** A pair of trochoid-like patterns, for $n = 1$, $\mu = 8$.

**Figure 3:** The amplitude of the edge wave, from the shoreline to the ocean, using equation (28) for $n = 2$; starting from the upper curve: $\lambda = 0 \cdot 2, 0 \cdot 1\dot{3}, 0 \cdot 0\dot{6}, 0$.

**Figure 4:** The amplitude of the edge wave (upper curves) in the presence of a sand bar (lower curves). The sand bar starts at (a) $Z = 1$; (b) $Z = 1 \cdot 75$. The curves (c) correspond to a uniformly sloping bottom, without a sand bar.

**Figure 5:** A three-dimensional plot of the surface wave (water in blue, bottom/beach in red), based on equation (29) with $n = 4$, $\mu = 4$.

**Figure 6:** The effective depth profile (lower curves) associated with a longshore current (upper curve) with $k = (\ell X_0)/\omega$, for $k = 0 \cdot 9$ and $k = 1 \cdot 4$.

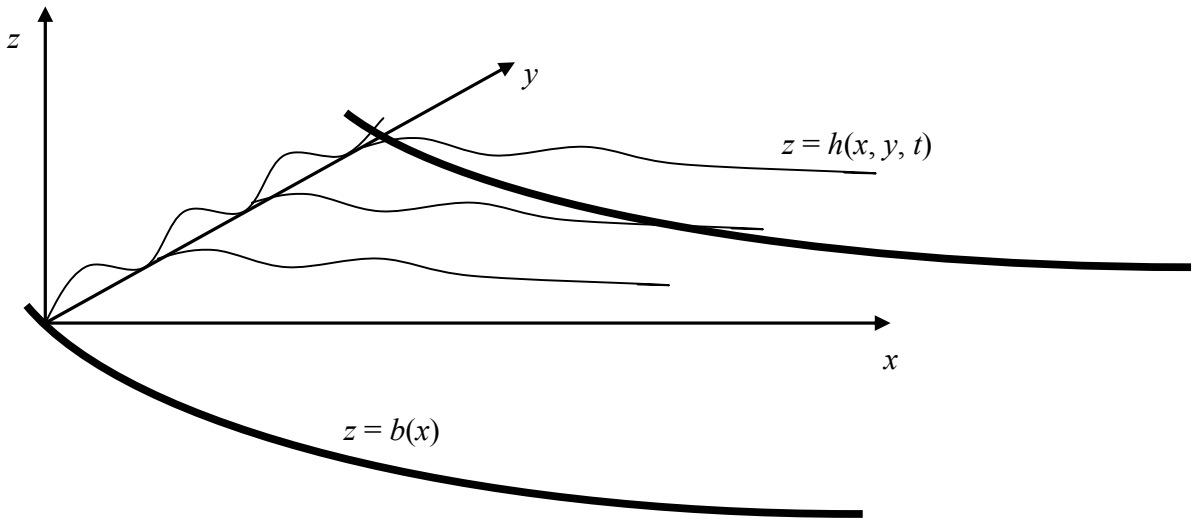

Figure 1

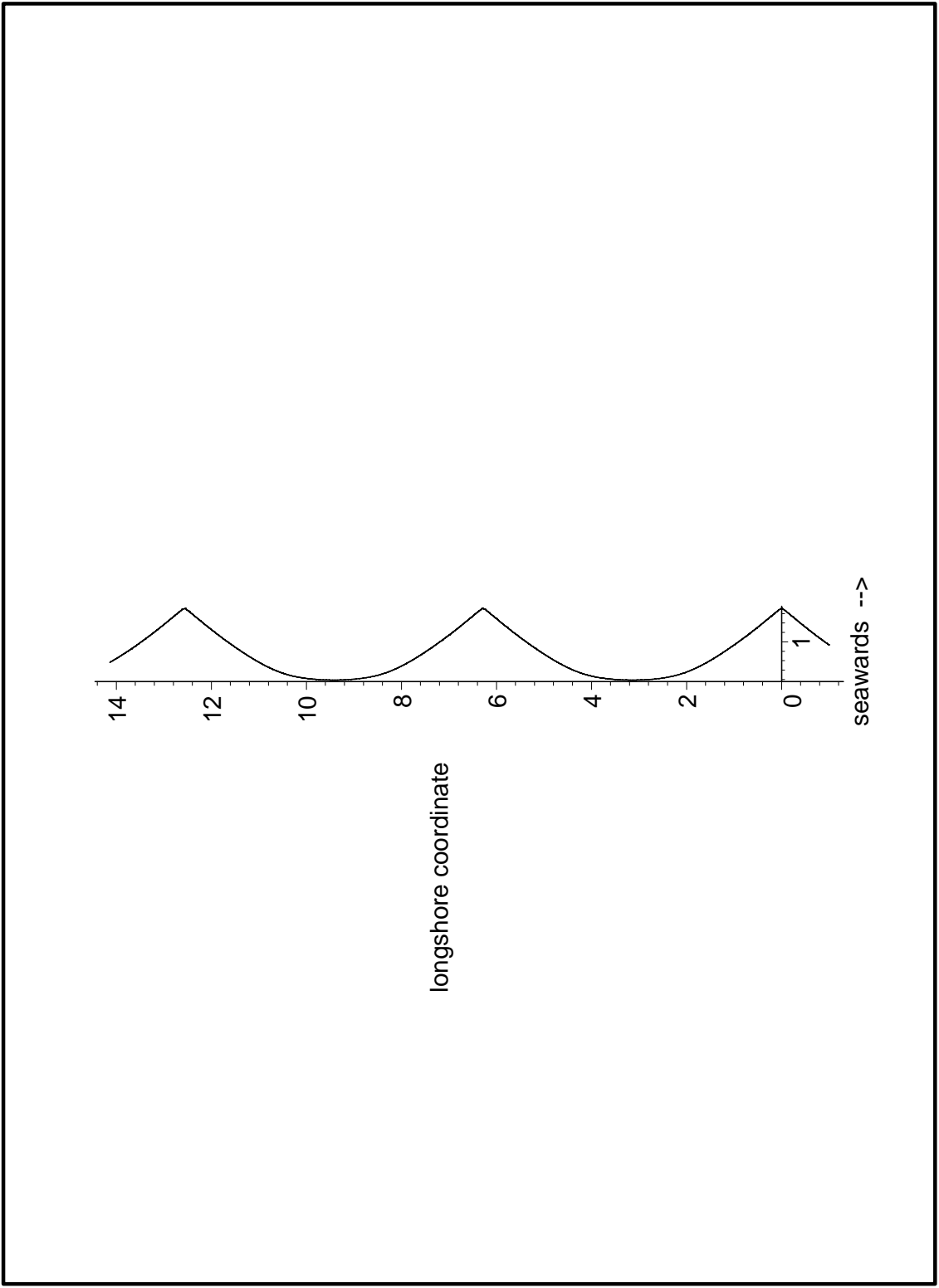

Figure 2a

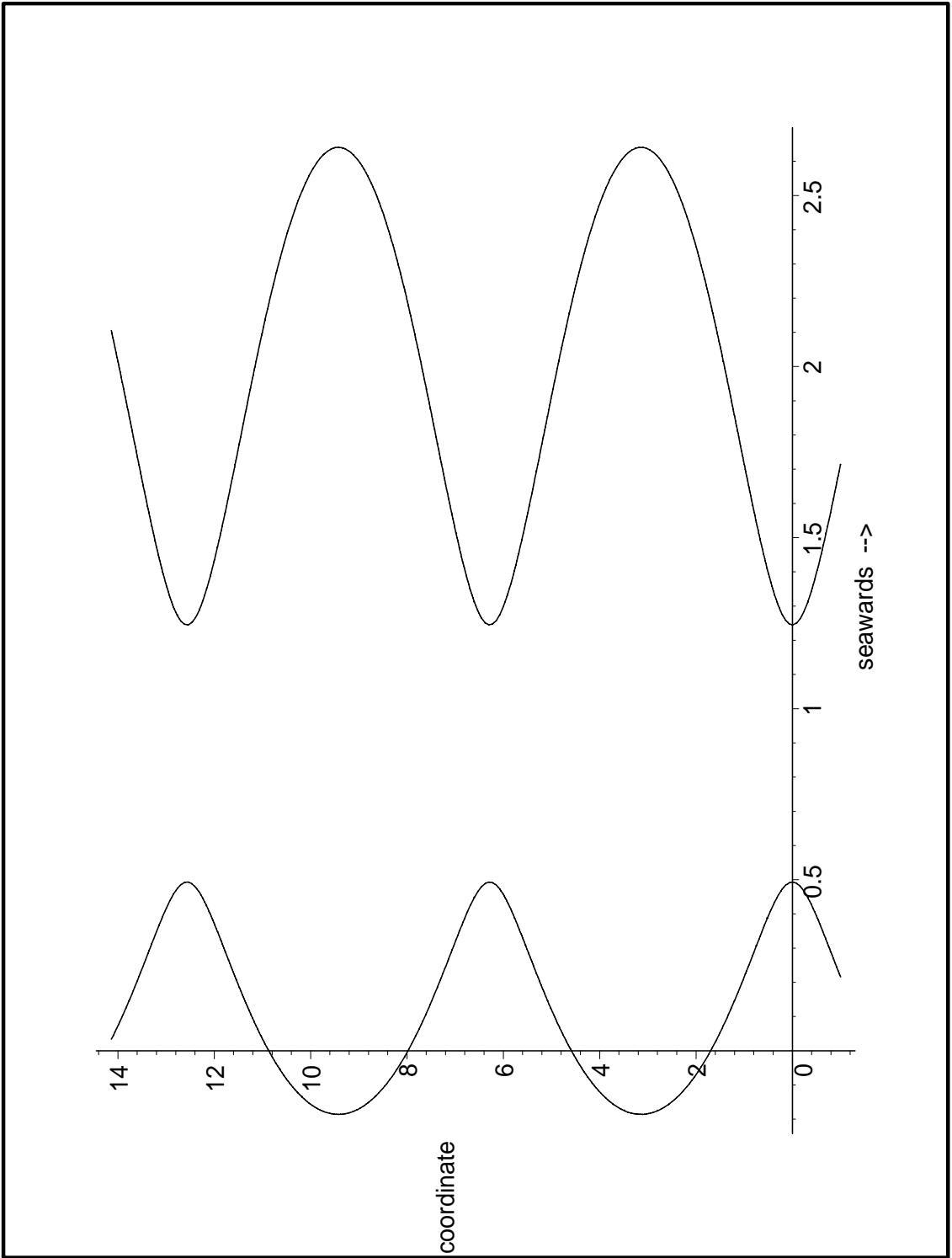

Figure 2b

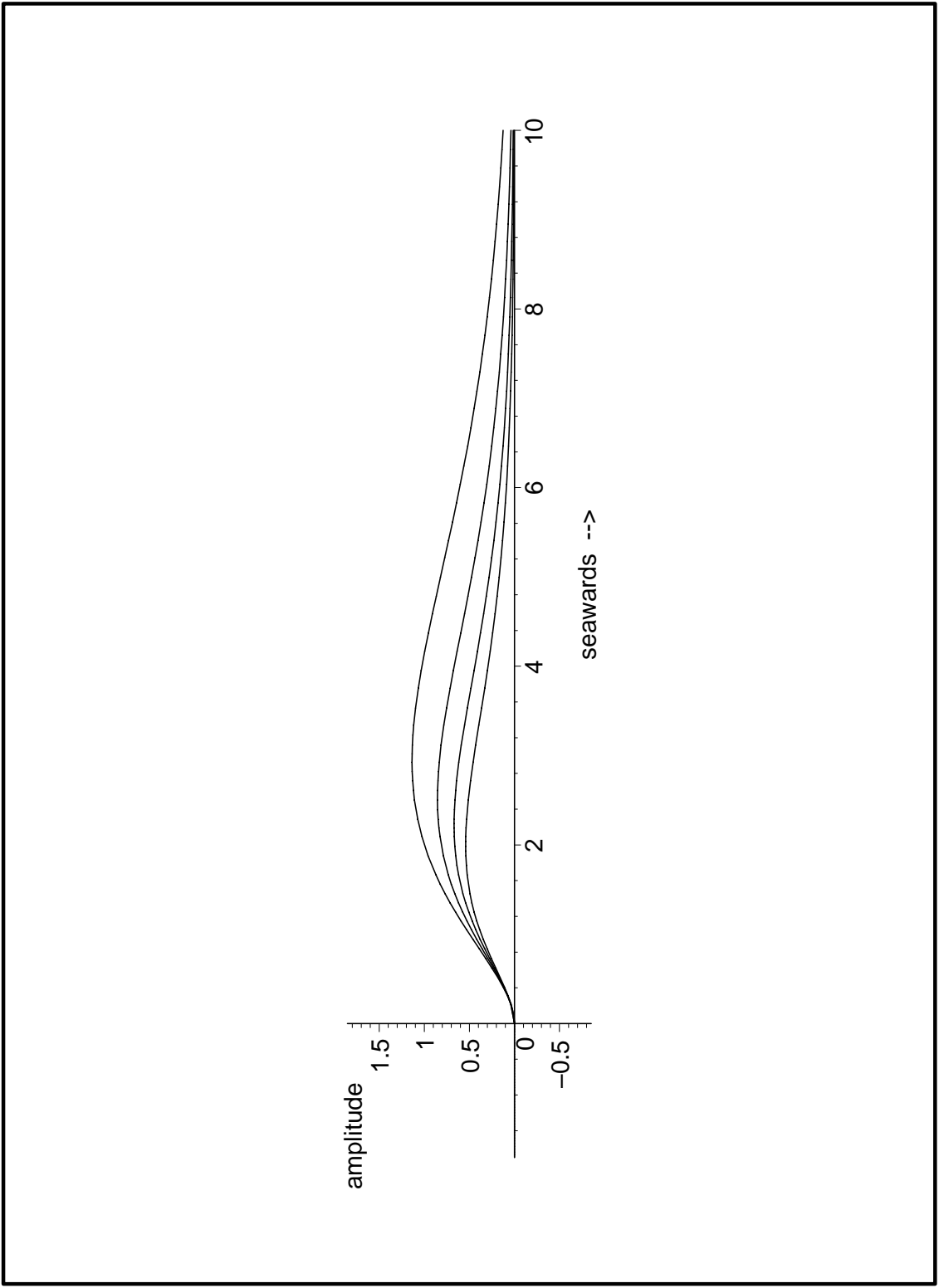

Figure 3

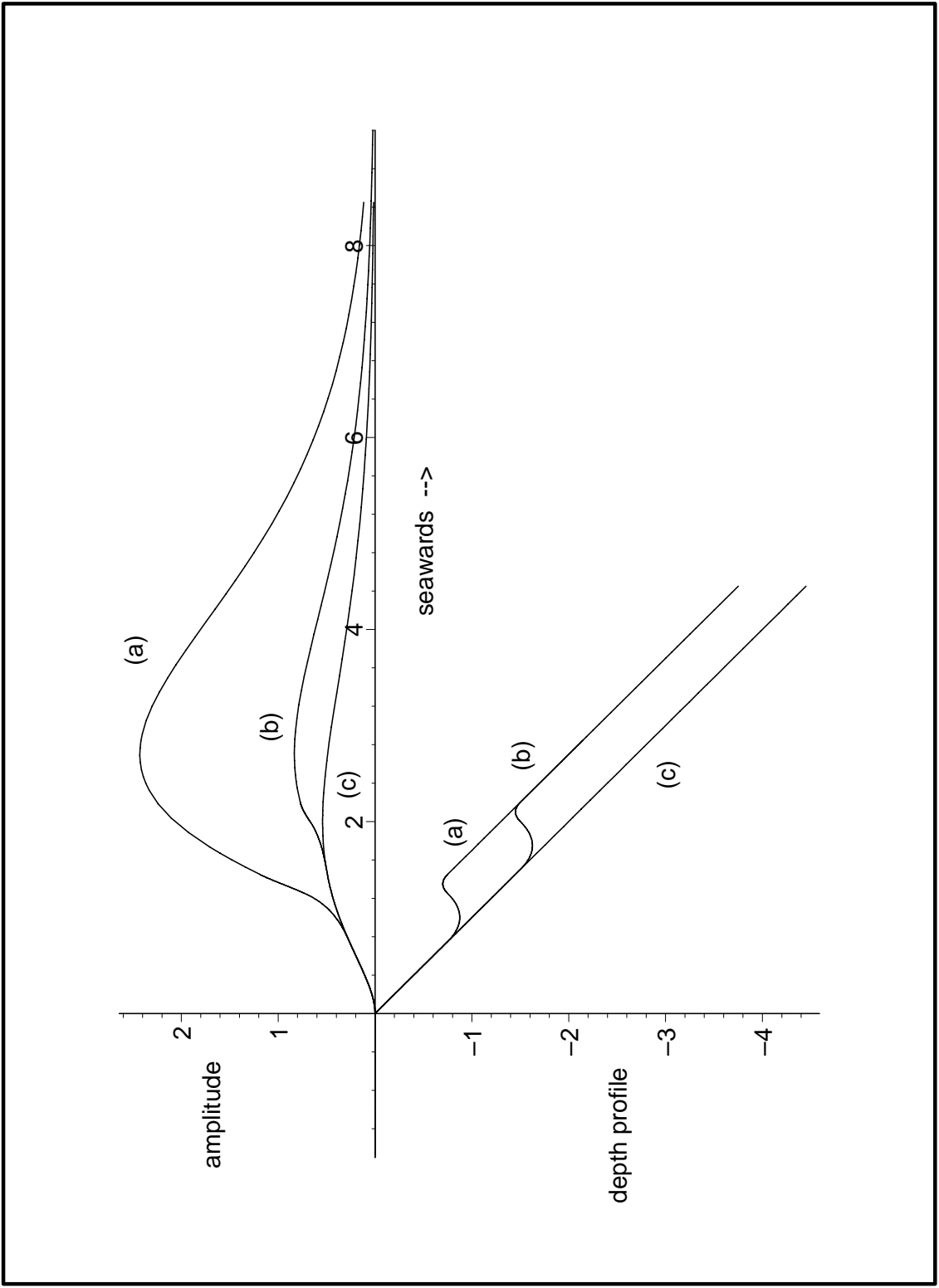

Figure 4

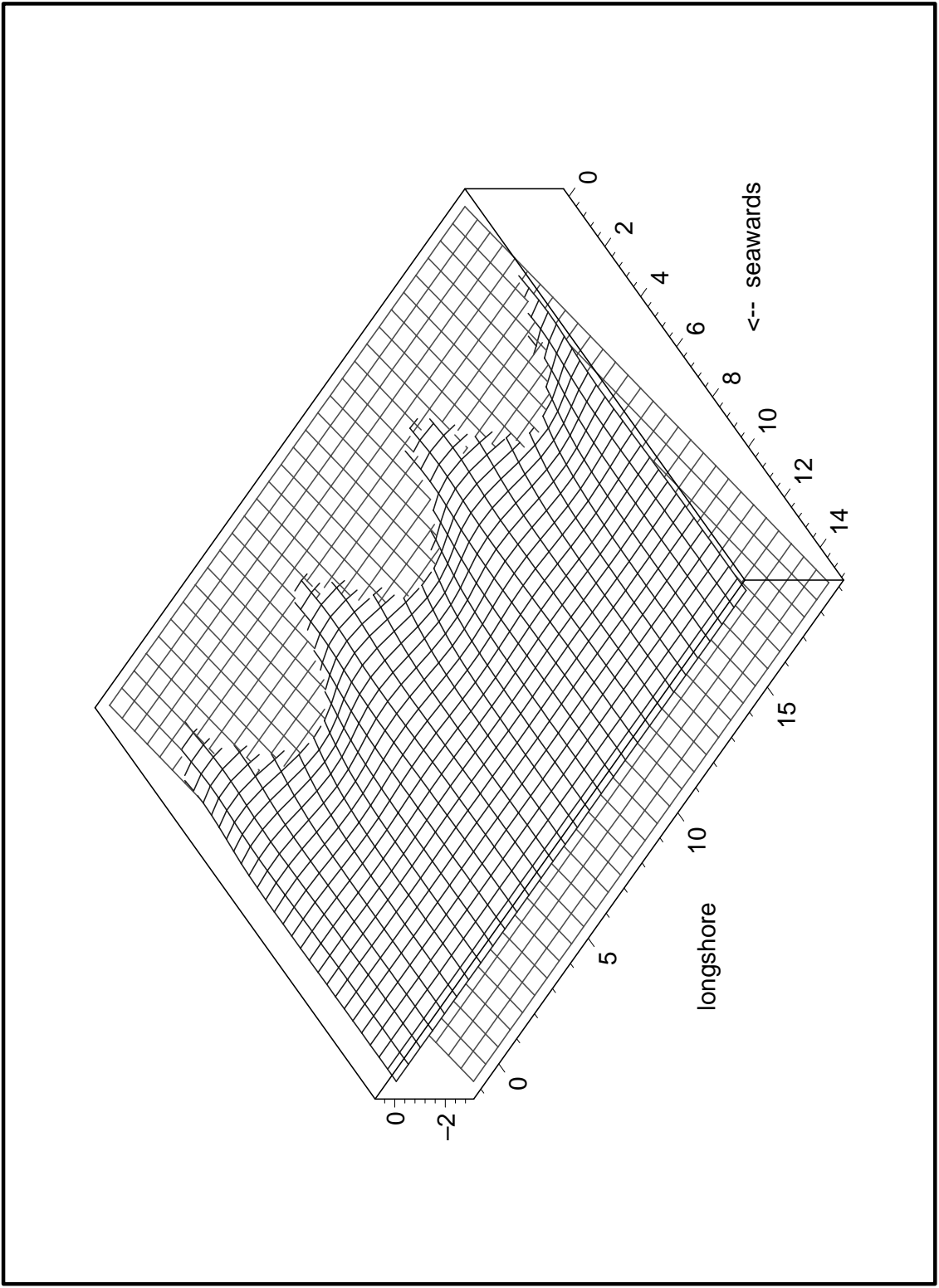

Figure 5

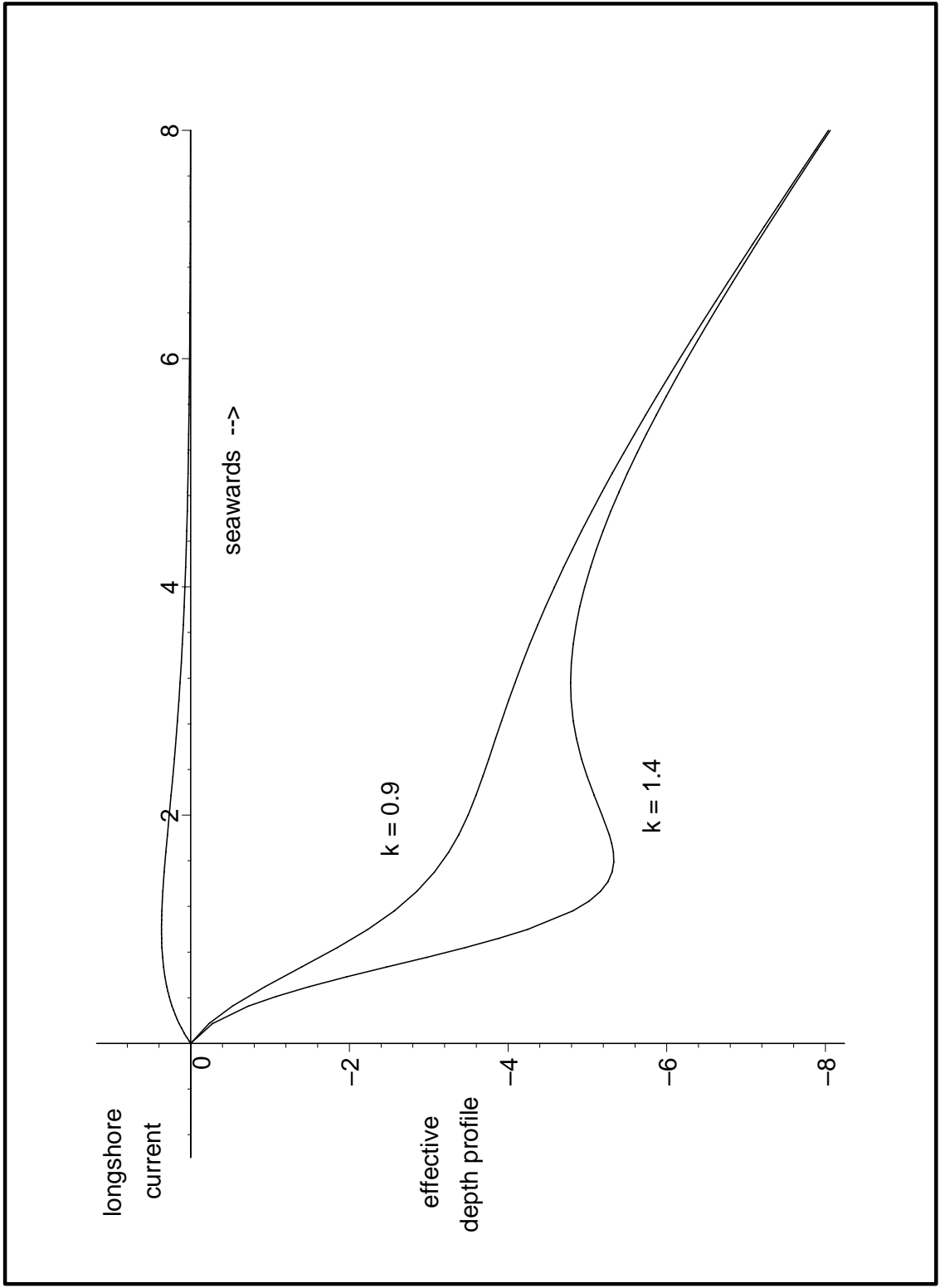

Figure 6